\documentclass[11pt]{article}
\usepackage[utf8]{inputenc}
\usepackage[T1]{fontenc}
\usepackage{amsmath,amsfonts,amssymb}
\usepackage{graphicx}
\usepackage{booktabs}
\usepackage{array}
\usepackage{placeins}
\usepackage[numbers,sort&compress]{natbib}
\usepackage[letterpaper,margin=1in]{geometry}
\usepackage{microtype}
\usepackage[font=small,labelfont=bf]{caption}
\usepackage{hyperref}
\usepackage{url}
\usepackage{xcolor}
\definecolor{linkblue}{RGB}{28,73,109}
\hypersetup{colorlinks=true,linkcolor=linkblue,citecolor=linkblue,urlcolor=linkblue,
    pdfauthor={Koutian Wu, You Liu, Yue Liu, Zhiping Li, Ergan Shang},
    pdftitle={Scientific Liquidity Agents: An Agent-Based Model of Knowledge Arbitrage in Scientific Communities}}
\graphicspath{{arxiv/figures/}}
\newcommand{\PrimaryPapers}{68}

\newcommand{\PrimarySolved}{134}

\newcommand{\ExtendedPapers}{274}

\newcommand{\ExtendedSolved}{360}

\title{\textbf{Scientific Liquidity Agents}:\\An Agent-Based Model of Knowledge Arbitrage\\in Scientific Communities\thanks{First version on Zenodo, December 8, 2025: \url{https://zenodo.org/records/17861999}}}
\author{\normalsize Koutian Wu\textsuperscript{1,2}, You Liu\textsuperscript{3}, Yue Liu\textsuperscript{3},\\
\normalsize Zhiping Li\textsuperscript{4}, Ergan Shang\textsuperscript{5}\\[0.5em]
\small\textsuperscript{1}The University of Texas at Austin\quad\textsuperscript{2}TaciteAI\\
\small\textsuperscript{3}Independent Researcher\\
\small\textsuperscript{4}Princeton University\quad\textsuperscript{5}Carnegie Mellon University\\[0.3em]
\small\texttt{k@tacite.ai}}

\date{}
\begin{document}
\maketitle
\begin{abstract}
Scientific knowledge gains new uses when ideas, methods, and results move between research contexts. Understanding this process requires a model of both specialized knowledge production and the intermediary activities that connect it. We introduce Scientific Liquidity Agents (SLA), an agent-based model of knowledge arbitrage in scientific communities. SLA represents domain specialists and epistemic arbitrageurs within a shared environment of research problems, researcher interactions, and scientific artifacts. Individual decisions connect to collective state through a belief--desire--intention architecture, a prerequisite-based problem graph, a dynamic interaction network, and a persistent publication registry. The framework makes researcher roles and knowledge-transfer frictions explicit, providing a process-oriented account of scientific production and exchange. Two exploratory simulations with 10 and 18 agents produce 68 and 274 registered papers. In the smaller run, arbitrageurs take part in 16 of the 26 contact pairs that join this community into one connected component, while topical similarity alone splits it into seven. The same run yields three similarity edges, 26 contact pairs, and two coauthorship pairs, so measured community structure depends on what counts as a link. Paper production is abundant, but no recorded replication succeeds, which places the bottleneck in verification rather than production. Solved-problem sets correspond to the prerequisite reachability of the disciplines represented in each community. SLA makes knowledge liquidity, the movement and reuse of artifacts across domains, a quantity that agentic science can simulate and measure.
\end{abstract}
\noindent\textbf{Keywords:} knowledge arbitrage; agent-based modeling; science of science; knowledge liquidity; scientific communities; epistemic intermediaries.

\section{Introduction}
Scientific progress depends on the circulation of existing knowledge as well as the production of new results. A method developed in one field may become an input to another, while a useful result may remain separated from potential users by disciplinary language, unfamiliar tools, or the effort required to adapt it. These movements connect individual research activities to the collective development of science~\citep{fortunato2018science}. They also create a role for researchers who work across disciplinary boundaries and help bring otherwise separated knowledge into contact.

We use \emph{knowledge arbitrage} to describe the pursuit of value from differences in where knowledge is available, understood, and potentially useful. An \emph{epistemic arbitrageur} occupies an intermediary position: its research profile emphasizes cross-domain activity, cooperation, and tools that can support the handling of unfamiliar work. \emph{Knowledge liquidity} describes the ease with which scientific artifacts can be discovered, understood, reproduced, and incorporated into subsequent research. Together, these concepts bring the scientific division of labor into view: specialists develop knowledge within domains, while intermediaries are positioned to connect knowledge across them.

Studying this process requires following multiple forms of change at once. Researchers choose activities and partners; interactions alter their contact histories; publication adds artifacts to a shared pool; and progress on one problem changes the prerequisites available for later work. A computational model can place these processes within a common environment and expose the relationships between local activity and collective state. The central question motivating this work is: \emph{How can knowledge arbitrage be represented as a process within an evolving scientific community, and what patterns of production, interaction, and cumulative progress arise in that environment?}

We introduce \textbf{Scientific Liquidity Agents (SLA)}, an agent-based model organized around this question. SLA combines domain specialists and epistemic arbitrageurs with a shared problem graph, interaction network, and publication registry. Researcher decisions use either an LLM-based structured-memory policy or an equation-based policy. Actions update the environment, whose evolving state informs subsequent decisions. This structure places specialized production and intermediary activity within the same continuing research process.

Our contributions are threefold:
\begin{enumerate}
\item \textbf{A process-oriented model of knowledge arbitrage.} We connect knowledge production, cross-domain interaction, and attempted reuse through explicit representations of heterogeneous researchers, transfer frictions, and intermediary roles. This formulation makes knowledge arbitrage a concrete object of computational investigation.
\item \textbf{An integrated multi-agent framework.} SLA links researcher decisions to a prerequisite-based problem graph, a dynamic interaction network, and a persistent publication registry, connecting individual activities with collective scientific state.
\item \textbf{An exploratory characterization of scientific-community dynamics.} Two simulations describe publication, cross-role contact, coauthorship, and problem progress. They exhibit extensive cross-role contact alongside sparse joint production, and correspondences between solved-problem sets, disciplinary coverage, and prerequisite structure.
\end{enumerate}

The paper develops the process model and its implementation, then uses the two simulation cases to examine how its different components evolve together. The resulting contribution is a computational account of knowledge arbitrage as a form of scientific division of labor.

\section{Related Work}
\paragraph{Scientific communities and interdisciplinary knowledge.}
Science-of-science research studies how individual choices, collaboration, and institutions shape collective scientific activity~\citep{fortunato2018science}. Agent-based models have examined paper generation and the formation of specialties~\citep{gilbert1997simulation}, communication and scientific learning~\citep{zollman2007communication}, and the relationship between individual research choices and aggregate interdisciplinarity~\citep{raimbault2020interdisciplinary}. SLA develops this tradition around a particular process: the coexistence of specialized knowledge production and cross-domain intermediary activity. It represents that process through persistent problems, contacts, and artifacts.

\paragraph{Language-model agents and social simulation.}
BDI architectures organize agent state into beliefs, desires, and intentions~\citep{rao1995bdi}. Generative Agents connects memory, reflection, and planning in a social environment~\citep{park2023generative}, while EconAgent studies LLM-driven activity in a macroeconomic setting~\citep{li2024econagent}. MatrAIx develops heterogeneous persona agents and interaction environments for simulated-user evaluation~\citep{li2026matraix}. SLA uses agent-based simulation to examine a domain-specific social process: how researcher roles and activities relate to the production and circulation of scientific knowledge.

\paragraph{Scientific ideation and research ecosystems.}
IdeaScientist retrieves mechanisms from analogous problems across fields to support grounded scientific ideation~\citep{liu2026ideascientist}. Academic Simulacra studies research-idea forecasting through multi-agent simulations~\citep{choi2025academicsimulacra}. SciUtopia models continuing academic ecosystems and investigates institutional mechanisms through counterfactual simulation worlds~\citep{jin2026sciutopia}. These studies address complementary scales of scientific activity, from individual proposals to institutional dynamics. SLA centers the intermediary role and the relationship among knowledge production, cross-domain contact, and prospective reuse. Its distinctive emphasis is the process through which a community's division of labor interacts with the structure of its accumulating knowledge.

\FloatBarrier

\section{Modeling Knowledge Arbitrage}
\label{sec:framework}
\subsection{Knowledge Production, Intermediation, and Use}
SLA models a scientific community as researchers acting on a shared collection of problems and artifacts. The process begins with specialized research: an agent works on a problem, contributes to its solution, and may publish a structured artifact. Publication makes that artifact available to later decisions. Collaboration creates contact between researchers, and replication provides an action through which an agent can attempt to use published work. These activities place knowledge creation and prospective circulation within one evolving system.

The knowledge-market analogy organizes three elements of the model: reusable assets, frictions in their movement, and intermediaries. Scientific knowledge is non-exclusive: one researcher's use can coexist with use by others. The model therefore focuses on access, effort, and use rather than ownership transfer or a transaction price. In the simulations reported here, an asset is a structured paper record with an identifier, authors, a publication step, and synthetic attributes. Problem completion, publication, and replication are state transitions in this computational environment.

Table~\ref{tab:concepts} maps the conceptual process to the simulator. Discoverability concerns which opportunities enter an agent's decisions. Tooling and replication costs concern the effort needed to handle a candidate artifact. Intermediary profiles combine cross-domain positioning with tool proficiency and cooperative tendencies. Knowledge liquidity provides the organizing question linking these elements: how readily can produced knowledge enter subsequent research?

\begin{table}[htbp]
\centering\small
\caption{Knowledge-arbitrage concepts and their representation in SLA.}
\label{tab:concepts}
\begin{tabular}{>{\raggedright\arraybackslash}p{0.19\linewidth}>{\raggedright\arraybackslash}p{0.35\linewidth}>{\raggedright\arraybackslash}p{0.37\linewidth}}
\toprule
Concept & Process interpretation & Model representation \\
\midrule
Knowledge production & Developing and communicating a result & Research actions, solved-problem records, and registered papers. \\
Intermediation & Connecting research across domains & Epistemic-arbitrageur profiles and cross-role collaboration opportunities. \\
Discovery & Encountering relevant people and artifacts & Ranked partner candidates, recent-paper candidates, and profile visibility. \\
Adaptation effort & Handling unfamiliar knowledge or tools & Tool proficiency, action costs, and domain-mismatch terms. \\
Attempted reuse & Trying to reproduce or use an artifact & Replication selection and downstream sampled outcomes. \\
Cumulative progress & Earlier results enabling further work & Problem prerequisites and persistent solver state. \\
\bottomrule
\end{tabular}
\end{table}

\subsection{Researcher Roles and Scientific Division of Labor}
Specialists and arbitrageurs instantiate the same agent class and share the action vocabulary. Specialists have named disciplinary assignments and topic profiles. Arbitrageurs have a generic cross-domain assignment, stronger initialized tool proficiency, and additional cooperation and helping parameters. This configuration represents the intermediary as a recognizable research role within the community. Table~\ref{tab:roleparams} specifies the role profiles used by the project runner.

\begin{table}[htbp]
\centering\small
\caption{Role initialization. $\mathrm{Beta}(a,b)$ denotes a beta draw; additive boosts are capped at one. The two roles use the same agent architecture.}
\label{tab:roleparams}
\begin{tabular}{>{\raggedright\arraybackslash}p{0.27\linewidth}>{\raggedright\arraybackslash}p{0.31\linewidth}>{\raggedright\arraybackslash}p{0.32\linewidth}}
\toprule
Attribute & Specialist & Arbitrageur \\
\midrule
Domain and topic & Named discipline and topic & \texttt{cross\_domain}; \texttt{interdisciplinary} \\
Skills & Programming/writing: $\mathrm{Beta}(2,2)$; mathematics: $\mathrm{Beta}(2,3)$; domain knowledge: $\mathrm{Beta}(3,2)$ & Same draws with $+0.2$ per skill \\
Cooperation & $\mathrm{Beta}(2,2)$ & Same draw with $+0.3$ \\
Tool proficiency & $\mathrm{Beta}(2,3)$ per tool & $\mathrm{Beta}(4,1)$ per tool \\
Visibility parameter & $\mathrm{Beta}(2,3)$ & $\mathrm{Beta}(4,2)$ \\
Incentive weights & Citations: $\mathrm{Beta}(3,2)$; workload: $\mathrm{Beta}(2,2)$; innovation: $\mathrm{Beta}(2,3)$ & Same weights, plus helping: $\mathrm{Beta}(4,1)$ \\
Other tendencies & Exploration: $\mathrm{Beta}(1.5,3)$; replication: $\mathrm{Beta}(1,3)$ & Same distributions \\
\bottomrule
\end{tabular}
\end{table}

Role parameters enter the model through different channels. Skills, tools, and cooperation affect synthetic action costs and success rules. Visibility is a profile field presented to the LLM; actual paper candidates come from the recent-publication registry. Helping is an initialized incentive field. The core runner supplies partner and paper candidates directly, while an additional network-information interface is available as an extension. Appendix~\ref{sec:eligibility} gives the exact candidate-selection rules, including the treatment of disciplinary labels.

\section{The SLA Framework}
\subsection{Individual Decisions and Shared State}
\begin{figure}[tbp]
\centering
\includegraphics[width=\linewidth]{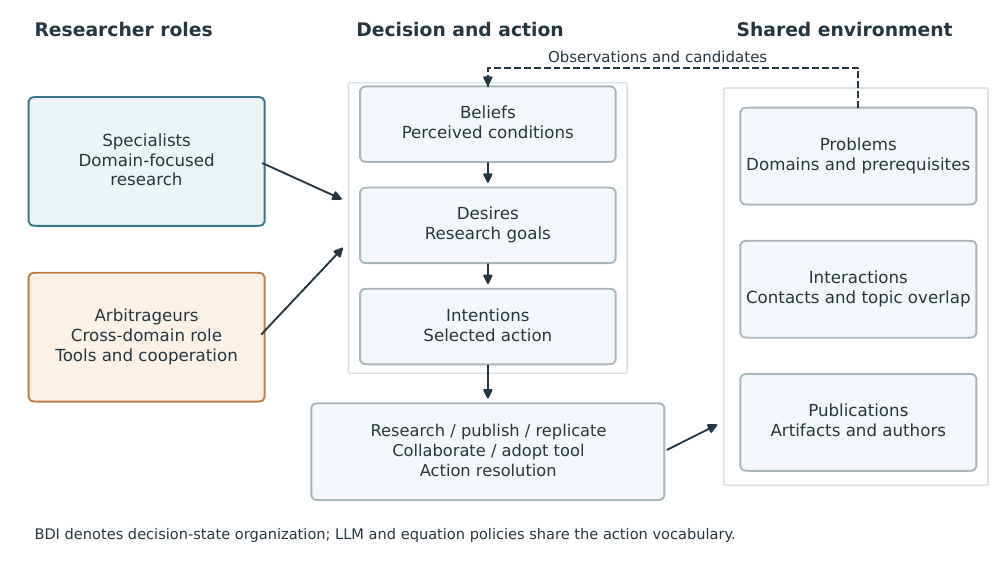}
\caption{SLA connects individual researcher decisions to shared scientific state. Role profiles inform the belief--desire--intention decision state. The five action types update problems, interactions, and publications, whose observations feed back into later decisions. The BDI sequence denotes conceptual state organization shared by the LLM and equation policies.}
\label{fig:framework_overview}
\end{figure}

Figure~\ref{fig:framework_overview} shows the decision--environment loop. Each researcher carries skills, tool proficiencies, behavioral tendencies, incentive weights, a knowledge inventory, and BDI state. Beliefs summarize perceived conditions, desires express goals, and intentions select actions. Shared scientific state consists of the problem graph, interaction state, and publication registry. A researcher's actions change the applicable shared records, which then inform subsequent decisions.

The LLM policy produces reasoning, structured-memory updates, a plan, and an action in a unified response. Numeric beliefs are updated separately. Parsed actions use the simulator's action vocabulary; invalid action names default to research, and missing or unparseable responses invoke equation-based intention selection. The equation policy uses numeric belief updates and rule-based goal and action selection. For example, the sentiment update combines persistence, observation, and social terms with coefficients 0.7, 0.3, and 0.2, followed by clipping to $[0,1]$.

Core actions are research, publication, replication, collaboration, and tool adoption. Each action has a synthetic effort cost modified by relevant profile attributes. The agent layer samples action success and updates local state. The environment then applies the corresponding operation: recording problem progress, deciding publication acceptance, resolving collaboration, or sampling a replication outcome. Appendix~\ref{sec:actionrule} specifies these rules. The sequence gives each activity a place in the ongoing community process, from individual choice to persistent shared output.

\subsection{Problems and Cumulative Research}
The problem graph contains research problems in mathematics, physics, chemistry, biology, and computer science. Each problem has a domain, difficulty, importance, prerequisites, and solver fields. Prerequisite relations represent cumulative dependence: progress on earlier problems can make later problems eligible for research. The graph therefore connects the temporal course of activity to the structure of the community's knowledge.

Problem queries use disciplinary assignments and prerequisite state to generate candidates. A successful research action can update solver fields and change later opportunities. Cross-domain dependencies make some local progress contingent on work in another field. The graph provides a common representation for studying both within-domain accumulation and interdependence across domains.

\subsection{Interaction and Artifact Circulation}
The interaction layer maintains time-decayed topic weights and persistent collaborator sets. Current similarity combines weighted Jaccard overlap with a contact bonus; a threshold defines current network edges. Partner candidates are ranked using similarity, role, domain, and contact history. In the ordinary collaboration branch, the agent selects the first returned candidate and the environment applies partner acceptance. Accepted interactions update the persistent contact record.

The publication registry stores accepted paper identifiers, authors, publication steps, and synthetic venue tiers. Registered publications enter the recent-paper pool used in later decisions. Replication selects among the latest 20 identifiers, with a preference for same-domain candidates when available. Its downstream success rule combines artifact clarity, agent programming skill, domain mismatch, and an optional friction-reduction term. Successful downstream draws record citations. These rules provide a stylized representation of artifact circulation and use within the simulated community.

\subsection{Observing the Process}
We examine four linked aspects of the model: production, interaction, problem progress, and replication. Production counts unique registered papers and their authorship. Interaction has three complementary representations: current similarity, persistent contact, and shared authorship. Problem progress follows solver state on the prerequisite graph. Replication records attempts and downstream outcomes under the simulator's rule.

The unit of each observable follows its role in the process. A contact is an opportunity for interaction; a coauthored paper is a joint product; a downstream replication outcome is a sampled use outcome. Keeping these units explicit allows their relationships to be studied together. Cross-role contact coverage uses $n_Sn_A$ possible pairs, where $n_S$ and $n_A$ are the numbers of specialists and arbitrageurs. Within-role coverage uses $\binom{n_t}{2}$. Publication credit is divided equally among registered authors, preserving the total number of papers.

The framework also includes an extension for artifact-level time to reuse, reuse frequency, replication throughput, and creation--reuse cost differences. These measures are implemented in a separate runner. The present cases use the production, interaction, problem, and replication records described above.

\section{Exploratory Simulation Setup}
\label{sec:protocol}
We study two existing SLA simulations. The primary community contains eight specialists and two arbitrageurs over 100 steps. The extended community contains sixteen specialists and two arbitrageurs over 200 steps. Table~\ref{tab:runs} summarizes their configurations and outputs. The cases provide two settings in which to examine how the framework's components operate together.

\begin{table}[htbp]
\centering\small
\caption{Simulation configurations and recorded outputs. Token counts are reported by the saved run summaries.}
\label{tab:runs}
\begin{tabular}{lrr}
\toprule
Measure & Primary run & Extended run \\
\midrule
Agents (specialists + arbitrageurs) & 8 + 2 & 16 + 2 \\
Simulation steps & 100 & 200 \\
Unique registered papers & 68 & 274 \\
Recorded replication outcomes & 0 & 6 \\
Successful replications & 0 & 0 \\
Solved problems / all records & 134 / 510 & 360 / 510 \\
Reported LLM tokens & 1,511,980 & 6,026,553 \\
\bottomrule
\end{tabular}

\end{table}

The primary specialists cover mathematics, chemistry, and biology; the extended community additionally includes physics and computer science specialists. Both cases use the same 510 problem definitions and 1,097 prerequisite edges. Both are documented as LLM-enabled; the primary protocol names Gemini 2.0 Flash and seed 42. Appendix~\ref{sec:provenance} describes the preserved inputs and computational reconstruction.

We count unique paper identifiers, derive interaction graphs from saved records, and compare solved-problem sets with prerequisite reachability. Starting with an empty set, the reachability calculation repeatedly adds problems whose prerequisites are already present. A domain-restricted calculation considers only domains represented by specialists in the corresponding roster. This static construction supplies a structural description against which to examine the simulated endpoints. The analysis characterizes each community and compares its observable patterns across these two configurations.

\section{Production, Interaction, and Cumulative Progress}
\label{sec:characterization}
\subsection{Scientific Production in the Two Communities}
\begin{figure}[htbp]
\centering
\includegraphics[width=\linewidth]{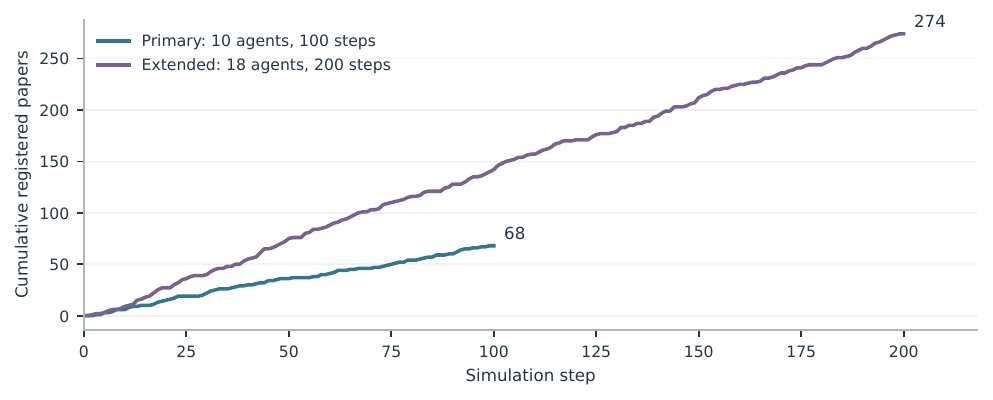}
\caption{Accumulation of unique registered papers in the two communities. Curves use the archived step sequence and publication registry, ending at 68 and 274 papers. Each legend identifies the community size and horizon.}
\label{fig:productiontime}
\end{figure}

The primary and extended simulations produce \PrimaryPapers{} and \ExtendedPapers{} unique registered papers, respectively. The primary registry contains two joint papers and the extended registry five. Figure~\ref{fig:productiontime} follows their accumulation over simulation steps; cumulative registry counts agree with all retained snapshots. These artifacts form the shared publication pool that later researcher decisions can encounter.

Both roles contribute registered output. Fractional authorship credit assigns 58 papers to specialists and 10 to arbitrageurs in the primary community, and 246.5 and 27.5 in the extended community. Per-agent averages are 7.25 and 5.00 in the primary case, and 15.41 and 13.75 in the extended case. This allocation locates each role within the community's production activity; the community sizes, disciplinary composition, and horizons are given in Table~\ref{tab:runs}.

Individual output also varies within roles. The two primary arbitrageurs receive fractional credits of 1 and 9; their extended-case counterparts receive 12.5 and 15. Figure~\ref{fig:individualoutput} preserves the individual distributions, complementing the role averages. Token trajectories and venue distributions appear in Figures~\ref{fig:tokentime} and~\ref{fig:venues}.

\subsection{Researcher Activity and Division of Labor}
\begin{figure}[htbp]
\centering
\includegraphics[width=\linewidth]{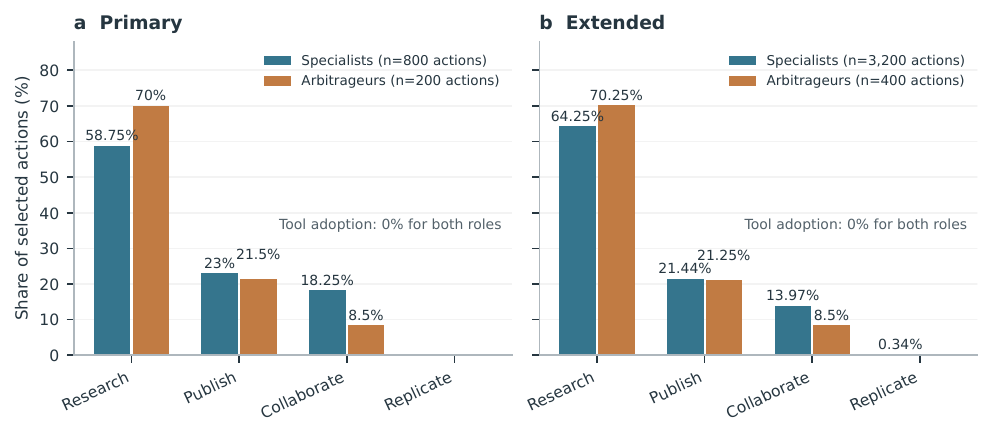}
\caption{Selected-action allocation by role. Percentages use 800/200 specialist/arbitrageur actions in the primary community and 3,200/400 in the extended community. The extended specialists' 0.34\% replication share is labelled explicitly. Tool adoption is zero in all four groups. These are action-choice proportions.}
\label{fig:actionallocation}
\end{figure}

Figure~\ref{fig:actionallocation} shows selected-action composition using each role's total agent-actions as its denominator. Primary specialists allocate 58.75\% to research, 23.00\% to publication, and 18.25\% to collaboration; arbitrageurs allocate 70.00\%, 21.50\%, and 8.50\%. The extended community has the same ordering of activities within each role. Its eleven replication selections occur among specialists, accounting for 0.34\% of their actions. Neither community selects tool adoption.

These activity profiles connect the configured division of labor to the choices made within the simulations. Arbitrageurs participate in direct research and publication as well as collaboration. The selected-action mix describes this participation alongside the contact and output patterns examined below. Appendix~\ref{sec:roles} gives the exact percentages and denominators.

\subsection{Cross-Role Contact and Joint Production}
\begin{figure}[htbp]
\centering
\includegraphics[width=\linewidth]{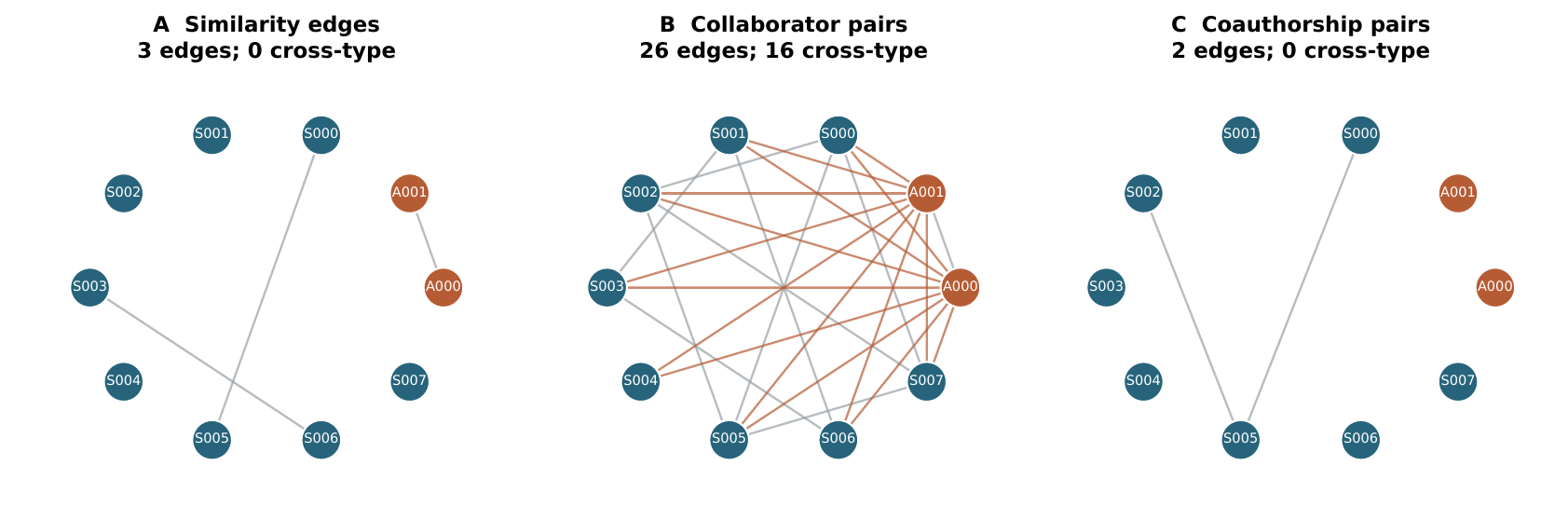}
\caption{Three network views of the primary community, with researcher positions held fixed across panels. Blue nodes are specialists and orange nodes are arbitrageurs. Panel A shows final similarity edges, panel B shows unique recorded collaborator pairs, and panel C connects coauthors of registered papers. Orange edges connect different roles. Both arbitrageurs contact every specialist, while joint authorship is concentrated in two specialist pairs.}
\label{fig:networks}
\end{figure}

Persistent contact reaches every specialist--arbitrageur pair: 16 of 16 pairs in the primary simulation and 32 of 32 in the extended simulation. Joint authorship is more selective. No cross-role pair shares a registered paper in the primary case, while three such pairs do in the extended case. Neither final similarity graph contains a cross-role edge. Figure~\ref{fig:networks} shows the three networks on the same node layout for the primary community. Figure~\ref{fig:networkcoverage} in the appendix compares normalized pair coverage across both communities.

These patterns locate contact and production at different stages of the modeled process. Persistent contact accumulates accepted encounters, current similarity reflects time-dependent topic overlap, and coauthorship identifies shared registered output. The partner-selection policy favors new cross-role contacts, giving a concrete route by which the contact network expands. Its saturated cross-role structure accompanies a much sparser structure of joint production.

\subsection{Disciplinary Coverage and Knowledge Dependencies}
The primary and extended cases finish with \PrimarySolved{} and \ExtendedSolved{} solved problems. Their solved sets exactly match the prerequisite closures restricted to the domains represented in the respective specialist rosters (Appendix Figure~\ref{fig:prerequisites}). The full-domain closure contains 360 problems. The remaining 150 belong to a 13-node cyclic component and its descendants. The primary roster introduces an additional domain restriction within this shared dependency structure.

\begin{figure}[htbp]
\centering
\includegraphics[width=\linewidth]{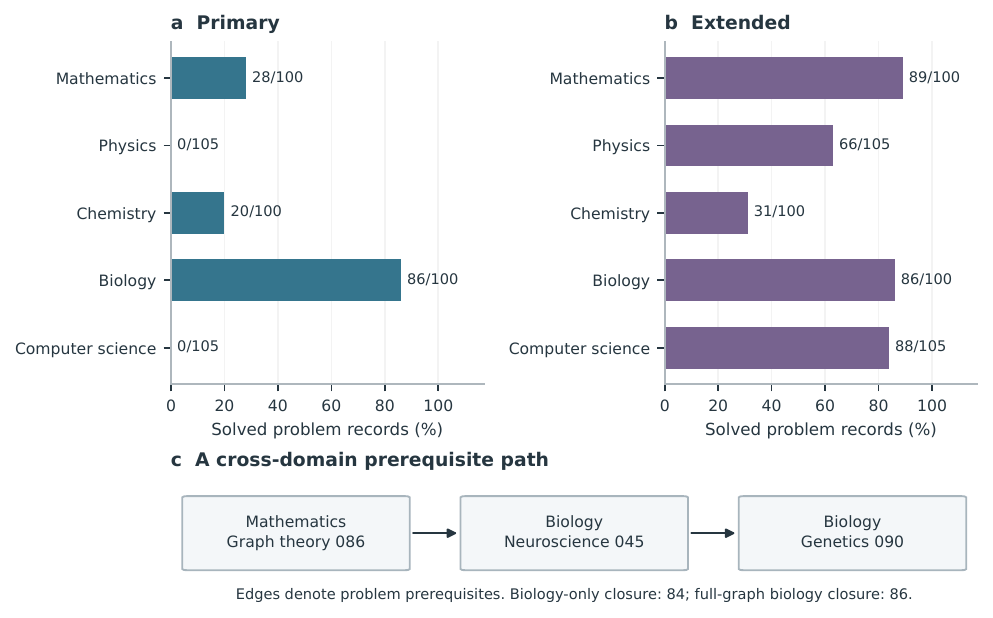}
\caption{Disciplinary progress and a cross-domain prerequisite path. Panels a and b show solved/total problem records for every discipline. Panel c follows one mathematics prerequisite into two biology problems. Its arrows are prerequisite relations in the problem graph. Biology-only closure reaches 84 problems, while the full-graph biology closure and both saved biology solved sets contain 86.}
\label{fig:disciplinaryprogress}
\end{figure}

Figure~\ref{fig:disciplinaryprogress} places disciplinary completion alongside a concrete prerequisite path. Biology illustrates how disciplinary and structural views complement one another. Both communities solve the same 86 biology problems. A closure restricted to biology alone reaches 84; the additional two problems require progress through a mathematics prerequisite. This explicit dependency links a local disciplinary outcome to work elsewhere in the knowledge graph. Appendix~\ref{sec:prerequisites} gives the problem identifiers and closure construction.

Equal endpoints also coexist with different trajectories. Among the 86 shared biology solutions, 84 solve steps and 57 paper-association lists differ between the two cases. Endpoint reachability describes the common structural destination, while the saved histories describe different temporal paths through it. These observations connect community composition, knowledge dependence, and the timing of cumulative research.

\FloatBarrier
\subsection{Replication Within the Activity Mix}
Replication is infrequent in these simulations. The primary summary contains no replication selection. The extended summary contains 11 selections and six agent-layer successes, together with six downstream records: one self-target and five other-agent targets. All six downstream outcomes are unsuccessful, and both final citation graphs are empty.

Within the modeled production--interaction--use process, the two cases therefore concentrate activity in artifact creation and researcher contact. Replication supplies a further, separately resolved stage. Its outcome depends on synthetic clarity, skill, and domain compatibility, as specified in Appendix~\ref{sec:replicationrule}. The cases give a concrete example of a scientific community with an expanding artifact pool and broad cross-role contact, yet little recorded activity at the downstream replication stage.

\section{Discussion}
\subsection{Knowledge Arbitrage as Scientific Division of Labor}
SLA makes knowledge arbitrage a question about the organization of research activity. Specialists and intermediaries participate in a shared environment, contribute artifacts, form contacts, and act within a common dependency structure. The intermediary role is represented explicitly, so its place in the community can be examined alongside specialized production. This perspective connects the potential usefulness of knowledge across domains with the researchers and activities through which that knowledge may circulate.

The exploratory cases illustrate this perspective in three ways. First, both roles participate in scientific production. Second, arbitrageurs connect the community: they take part in 16 of the 26 primary-run contact pairs, which join agents that topical similarity alone separates into seven components. Third, problem progress follows a shared prerequisite structure while its temporal paths differ. Together, these observations show why knowledge arbitrage is naturally studied through the joint evolution of people, artifacts, and problems.

\subsection{From Contact to Cumulative Research}
The simulations give contact, production, and replication distinct places in the community's activity. Contact links researchers; publication adds a shared artifact; replication attempts its use. Contact is abundant and successful replication is absent, so the transitions between these stages determine how much knowledge circulates. For a process model, the important object is how opportunities enter later activity over time.

The dependency graph adds another dimension to this account. The two biology problems requiring a mathematics prerequisite illustrate a structural channel through which progress in one field supports possibilities in another. The different biology trajectories show how communities can arrive at the same set of solutions through different histories. This combination makes both timing and dependency structure relevant to the study of knowledge circulation.

These observations motivate a focused research agenda around the allocation of effort among production, discovery, adaptation, and use. The role profiles, costs, and candidate rules provide concrete places to vary that allocation. Questions include when a community benefits from dedicated intermediaries, how information access interacts with adaptation effort, and how the structure of prerequisites changes the timing of cross-domain opportunities. The present framework organizes these questions around a common computational process.

\subsection{Model Scope and Extensions}
SLA represents papers as structured records, problem solutions as simulator state changes, and replication outcomes as explicit stochastic rules. This explicitness lets each stage of knowledge circulation be varied and measured on its own.

Richer artifact representations would allow adaptation to operate on scientific content directly. Connecting papers, datasets, or executable methods to recipient tasks would extend the same architecture from structured records to content-bearing transfers. Empirical calibration of researcher profiles and activity costs would support comparisons with particular scientific communities. The framework's modular separation of decisions, shared state, and action resolution provides a basis for such extensions.

\section{Conclusion}
We present Scientific Liquidity Agents, a framework that makes knowledge flow among LLM researcher agents observable and measurable. The archived runs yield three insights. First, cross-domain arbitrageurs act as bridges: they take part in 16 of the 26 contact pairs that join the primary-run community into one connected component; topical similarity alone splits the same community into seven. Second, the measured structure of a research community depends on what counts as a link: the same run yields three similarity edges, 26 contact pairs, and two coauthorship pairs. Third, the bottleneck lies in verification, not production: the runs register \PrimaryPapers{} and \ExtendedPapers{} papers, and replication is the step that no agent completes successfully.

IdeaScientist~\citep{liu2026ideascientist} shows that agents can draw mechanisms from analogous fields to generate research ideas. SLA models what happens after generation: how artifacts move between domains and whether other agents verify and reuse them. Knowledge liquidity thus becomes a quantity that agentic science can simulate, measure, and target, alongside knowledge production.

\paragraph{AI assistance.}
Generative AI coding and writing assistants supported framework development, manuscript preparation, code inspection, and offline analysis. The authors are responsible for the manuscript and its interpretation.

\paragraph{Data and code availability.}
The accompanying version directory contains manuscript sources, frozen simulation records, selected source snapshots, and scripts for reproducing the reported descriptive analysis. Input provenance and SHA-256 hashes are recorded in \path{arxiv/manifest.json}; build and analysis commands are provided in \path{README.txt}.

\clearpage
\bibliographystyle{plainnat}
\bibliography{references}
\clearpage
\appendix
\section{Supplementary Views of Research Activity}
\label{sec:supplementary}
\begin{figure}[htbp]
\centering
\includegraphics[width=\linewidth]{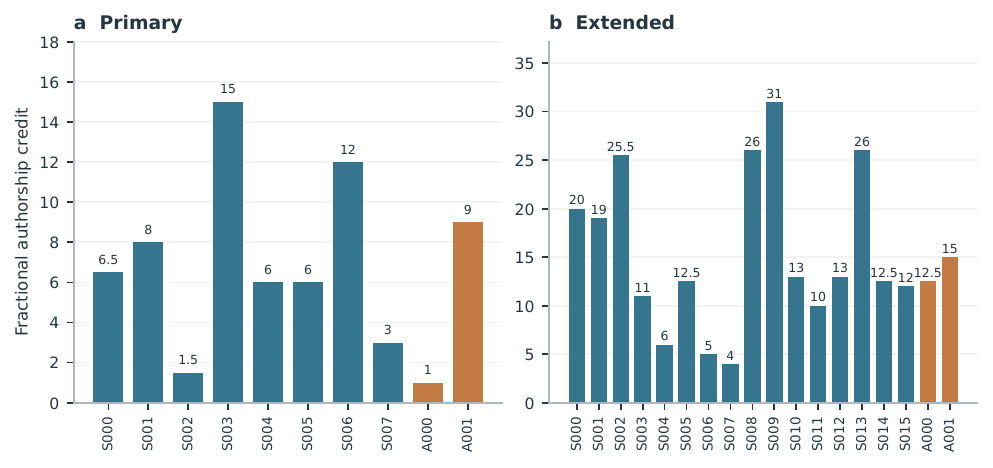}
\caption{Individual fractional authorship credit. The primary arbitrageurs contribute 1 and 9 credits, while the extended arbitrageurs contribute 12.5 and 15. Credits across all agents sum to 68 and 274.}
\label{fig:individualoutput}
\end{figure}

Figure~\ref{fig:individualoutput} shows output at the individual level. Blue bars denote specialists and orange bars arbitrageurs; each agent retains its archived numerical identifier. Fractional credit allocates joint papers equally across their recorded authors.

\begin{figure}[htbp]
\centering
\includegraphics[width=\linewidth]{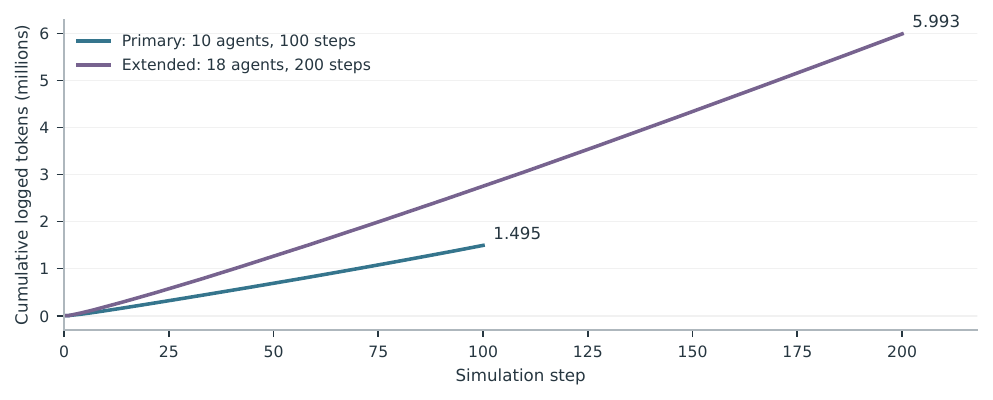}
\caption{Logged cumulative token use. The plot retains the saved time-series values, displayed in millions. Community sizes and horizons are identified in the legend.}
\label{fig:tokentime}
\end{figure}

Figure~\ref{fig:tokentime} retains the logged token trajectories. Their last values are 1,494,803 and 5,992,945, compared with final-summary totals of 1,511,980 and 6,026,553. The differences are 17,177 and 33,608 tokens. Curves use the logged series; Table~\ref{tab:runs} uses the final summaries.

\begin{figure}[htbp]
\centering
\includegraphics[width=\linewidth]{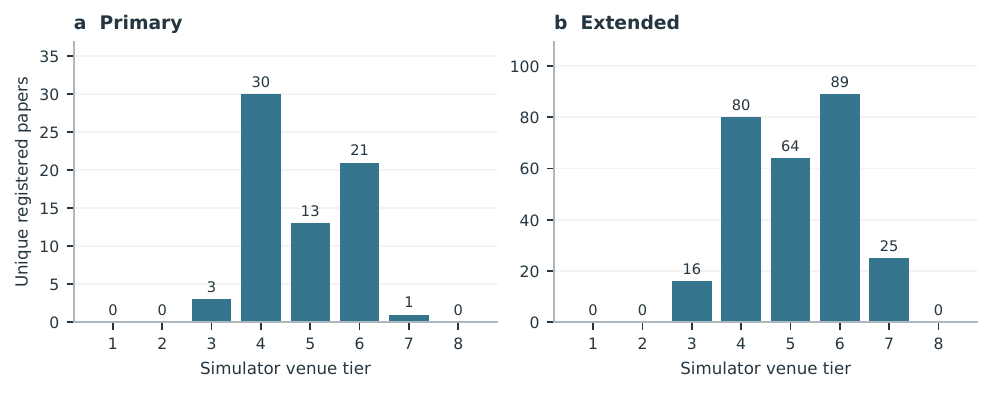}
\caption{Unique registered papers by simulator venue tier. Tier 1 is the most selective configured venue; tier 8 is the preprint category. Counts describe the publication environment's output distribution.}
\label{fig:venues}
\end{figure}

\clearpage
Figure~\ref{fig:venues} describes the publication subsystem through its configured venue tiers. Primary papers fall in tiers 3--7 with counts 3, 30, 13, 21, and 1; extended counts are 16, 80, 64, 89, and 25. These are categories assigned within the simulator.

\begin{figure}[htbp]
\centering
\includegraphics[width=\linewidth]{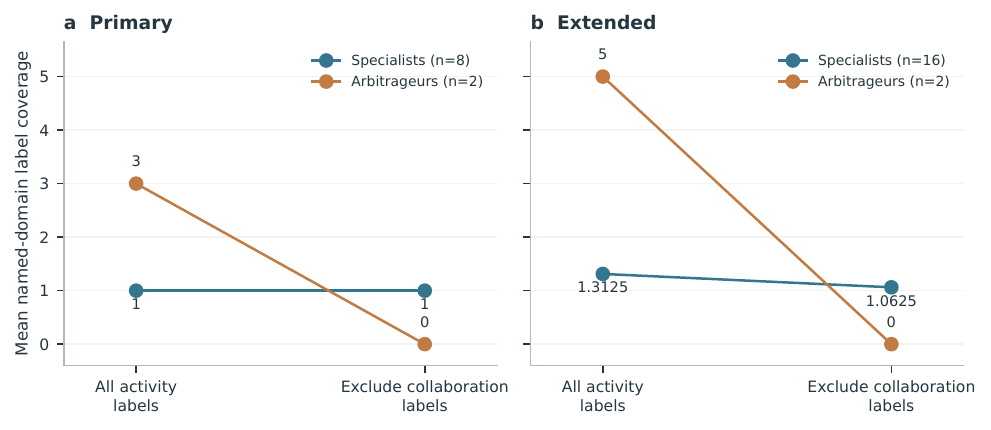}
\caption{Mean named-domain coverage in recorded activity labels under two counting conventions. Lines join summaries of the same agents after recoding; they are not temporal trajectories. Values describe labels rather than independent measurements of expertise or knowledge transfer.}
\label{fig:labelcoverage}
\end{figure}

Figure~\ref{fig:labelcoverage} reports named-domain coverage in activity-topic keys. It includes both the original label convention and the convention excluding keys ending in \texttt{\_collab}. Specialist means are 1.0/1.0 in the primary case and 1.3125/1.0625 in the extended case. Arbitrageur means change from 3 to 0 and from 5 to 0. Generic \texttt{cross\_domain} and \texttt{interdisciplinary} labels remain; the plotted quantity counts named-domain mentions.

\clearpage

\section{Simulation Rules and Additional Results}
\subsection{Role Composition and Activity}
\label{sec:roles}
Table~\ref{tab:composition} gives the specialist composition of each community. Each also includes two arbitrageurs assigned the generic \texttt{cross\_domain} label. Table~\ref{tab:actions} reports selected actions as a percentage of the corresponding role's agent-actions.
\begin{table}[htbp]
\centering\small
\caption{Specialist domain composition.}
\label{tab:composition}
\begin{tabular}{lrrrrr}
\toprule
Run & Math. & Physics & Chemistry & Biology & CS \\
\midrule
Primary & 4 & 0 & 1 & 3 & 0 \\
Extended & 4 & 2 & 2 & 2 & 6 \\
\bottomrule
\end{tabular}

\end{table}
\begin{table}[htbp]
\centering\small
\caption{Role-specific action distributions. S denotes specialists and A arbitrageurs.}
\label{tab:actions}
\begin{tabular}{lrrrrr}
\toprule
Run / role & Agent-actions & Research & Publish & Collaborate & Replicate \\
\midrule
Primary / S & 800 & 58.75\% & 23.00\% & 18.25\% & 0.00\% \\
Primary / A & 200 & 70.00\% & 21.50\% & 8.50\% & 0.00\% \\
Extended / S & 3,200 & 64.25\% & 21.44\% & 13.97\% & 0.34\% \\
Extended / A & 400 & 70.25\% & 21.25\% & 8.50\% & 0.00\% \\
\bottomrule
\end{tabular}

\end{table}

The primary final similarity graph has three edges and seven components. Its persistent-contact graph contains 26 pairs, and its coauthorship graph contains two pairs. Extended-case edge counts are four, 61, and five, respectively. Figure~\ref{fig:labelcoverage} gives the named-domain label summaries.

\begin{figure}[htbp]
\centering
\includegraphics[width=\linewidth]{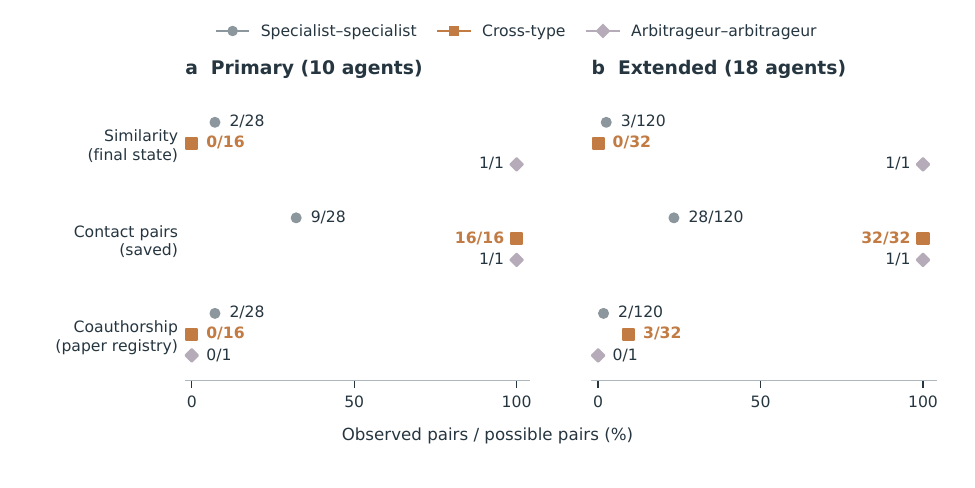}
\caption{Normalized pair coverage across both communities, complementing the primary-community network structure in Figure~\ref{fig:networks}. Cross-role coverage uses $n_Sn_A$ possible pairs; within-role coverage uses $\binom{n_t}{2}$. Similarity, persistent contact, and coauthorship retain their respective definitions.}
\label{fig:networkcoverage}
\end{figure}

\subsection{Problem Candidates and Partner Selection}
\label{sec:eligibility}
The selected source snapshot assigns arbitrageurs the domain \texttt{cross\_domain}. Research requests pass the agent's domain to \texttt{get\_solvable\_problems} and, if needed, \texttt{get\_frontier\_problems}. Both functions filter candidates by exact domain equality. Saved problems have named domains, so the generic cross-domain label returns no candidates under these query rules. The problem results in the main text are consequently analyzed at the community level. All saved non-null solver identifiers belong to specialists.

The included deterministic query fixture contains five prerequisite-free roots, one per discipline, and one mathematics problem depending on the mathematics root. Each named-domain query returns its root; the generic label returns zero; a query with no domain restriction returns all five roots. Once the mathematics prerequisite is satisfied, its dependent problem becomes eligible. The 18 fixed query cases describe the snapshot's candidate-selection behavior.

Collaboration candidates exclude existing collaborators. Cross-role candidates receive a score floor of 0.3, with an additional multiplier of 1.5 for arbitrageur requesters. Distinct domains receive a multiplier of 1.2. The ordinary collaboration branch selects the first ranked candidate, applies acceptance, and then records contact. The source also contains a separate bridging branch; neither saved action summary selects it. These rules explain how the simulator supplies opportunities for new cross-role contact.

\subsection{Action Resolution}
\label{sec:actionrule}
Base action costs are 10 for research, 5 for publication, 8 for replication, 6 for collaboration, and 3 for tool adoption. Relevant skill, tool, or cooperation factors modify these synthetic units. Given effective cost $C_i(a)$ and an action-specific bonus $b_i(a)$, the agent layer samples success with probability
\begin{equation}
 p_i(a)=\min\left(0.95,\frac{1}{1+0.1C_i(a)}+b_i(a)\right).
\end{equation}
Local memory and inventory updates occur at this layer. The environment then performs action-specific resolution. Publication has a venue decision, collaboration has partner acceptance, and replication has a downstream sampled outcome. The resulting records retain the meanings of their respective layers.

\subsection{Replication Rule}
\label{sec:replicationrule}
The replication branch draws candidates from the latest 20 paper identifiers. If same-domain candidates exist, it selects that subset with probability 0.7; otherwise it uses the full candidate set. A target is sampled uniformly from the selected set, including the agent's own papers when present. Given paper clarity $c_p$, programming skill $s_i$, optional friction reduction $f_i$, and domains $d_i,d_p$, the effective downstream success probability is
\begin{equation}
 q_{ip}=\max\!\left[0,\min\!\left(0.95,c_ps_i+f_i-0.2\,\mathbf{1}\{d_i\ne d_p\}\right)\right].
\end{equation}
The optional term is 0.3 times the agent's friction-reduction method when available, and zero otherwise. A successful draw records a citation. Clarity and the outcome are internal model quantities. Local inventory updates can precede this downstream draw and can refer to a separately selected target.

The six extended-case records all have unsuccessful downstream outcomes. Under a reference calculation with zero friction reduction and unit additional cost multipliers, their success probabilities are approximately $[0.0876,0,0.1411,0.1530,0.1818,0.3206]$. These values describe the six fixed targets under the reference rule. The saved observed outcomes remain the quantities reported in the main text.

\section{Problem Structure and Production Accounting}
\subsection{Static Prerequisite Closure}
\label{sec:prerequisites}
\begin{figure}[htbp]
\centering
\includegraphics[width=\linewidth]{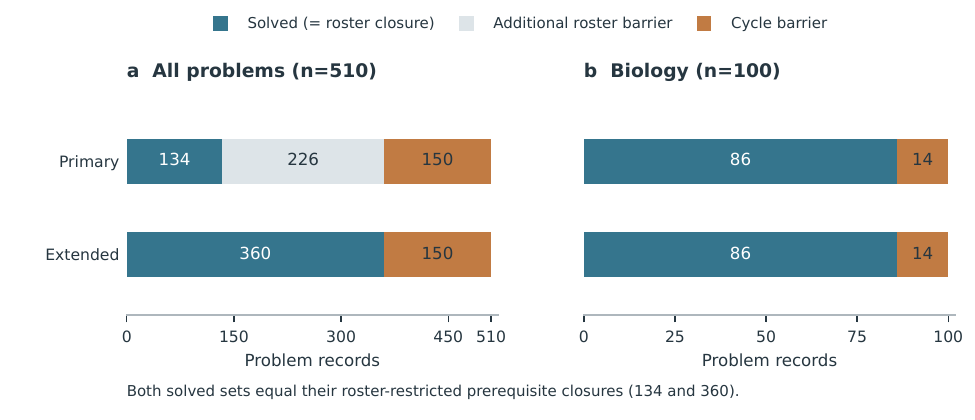}
\caption{Saved solved sets and empty-start prerequisite closure. The construction requires all prerequisites and uses no external seeds. Roster and cycle categories partition the problems outside each closure under this definition. Both biology solved sets contain 86 records.}
\label{fig:prerequisites}
\end{figure}

The two frozen dictionaries have identical problem identifiers, domains, difficulties, importance values, and prerequisite lists after removing solver, solve-step, and paper-association fields. Define $D$ as a set of candidate domains, $\operatorname{pre}(v)$ as problem $v$'s prerequisites, and $d(v)$ as its domain. Empty-start closure is the fixed point of
\begin{equation}
 C_0=\varnothing,\qquad
 C_{k+1}=C_k\cup\{v:d(v)\in D,\ \operatorname{pre}(v)\subseteq C_k\}.
\end{equation}
Across all five domains, the fixed point contains 360 problems. A 13-node cyclic strongly connected component and its descendants account for the remaining 150. Restriction to the primary roster's domains yields 134; the extended roster covers all domains and yields 360. These sets coincide exactly with the saved solved sets. Figure~\ref{fig:prerequisites} displays the resulting partition.

The closure uses an all-prerequisites rule and no external prerequisite seeds. It describes graph reachability. The source's candidate queries check prerequisites, while its solve mutator updates problem state without a separate eligibility check. We use closure as a static structural comparison with the saved endpoints.

Both biology solved sets contain 86 problems; biology-only closure contains 84. The additional problem \path{biology_neuroscience_045} depends on \path{mathematics_graph_theory_086}, and \path{biology_genetics_090} depends on that biology problem. The other 14 biology problems descend from the cyclic component. Solve steps differ for 84 of the 86 shared solutions, and paper-association lists differ for 57.

\begin{table}[htbp]
\centering\small
\caption{Problem counts and saved solved sets across disciplines.}
\label{tab:domains}
\begin{tabular}{lrrr}
\toprule
Domain & Total & Primary solved & Extended solved \\
\midrule
Mathematics & 100 & 28 & 89 \\
Physics & 105 & 0 & 66 \\
Chemistry & 100 & 20 & 31 \\
Biology & 100 & 86 & 86 \\
Computer science & 105 & 0 & 88 \\
\midrule
Total & 510 & 134 & 360 \\
\bottomrule
\end{tabular}
\end{table}

\subsection{Publication Units}
\label{sec:publicationcounts}
Production counts use unique registered identifiers. Fractional credit for agent $i$ is
\begin{equation}
 P_i=\sum_{p\in\mathcal P}\frac{\mathbf{1}\{i\in A_p\}}{|A_p|},
 \qquad \sum_i P_i=|\mathcal P|,
\end{equation}
where $A_p$ is a paper's recorded author set. Primary credits total 58 for specialists and 10 for arbitrageurs; extended credits total 246.5 and 27.5. Each joint paper has two authors. The 68 and 274 unique papers therefore correspond to 70 and 279 registered author--paper incidences.

The implementation also maintains personal publication lists, appending after agent-layer success and after registered authorship. Their 284 and 960 entries reconcile as $214+70$ and $681+279$. Table~\ref{tab:accounting} gives the units at each recording layer. The LLM profile in the source uses personal-list length; the production analysis uses the unique publication registry.
\begin{table}[htbp]
\centering\small
\caption{Selected actions, intermediate successes, and registered output. Each row retains its own counting unit.}
\label{tab:accounting}
\begin{tabular}{llrr}
\toprule
Record & Counting unit & Primary & Extended \\
\midrule
\multicolumn{4}{l}{\textbf{Publication}} \\
Internal publication list & List entries & 284 & 960 \\
Selected publication actions & Agent-actions & 227 & 771 \\
Agent-layer successes & Successful flags & 214 & 681 \\
Registered authorships & Author--paper pairs & 70 & 279 \\
Registered papers & Distinct paper IDs & 68 & 274 \\
\multicolumn{4}{l}{\textbf{Replication}} \\
Selected replication actions & Agent-actions & 0 & 11 \\
Agent-layer successes & Successful flags & 0 & 6 \\
Saved outcomes & Outcome records & 0 & 6 \\
Other-agent targets & Outcome records & 0 & 5 \\
Self-targets & Outcome records & 0 & 1 \\
Successful outcomes & Outcome records & 0 & 0 \\
Citations & Graph edges & 0 & 0 \\
\bottomrule
\end{tabular}

\end{table}

The complete problem dictionary contains 510 records. A legacy domain-indexed summary covers 500 and reports 352 extended-case solutions. Including three additional solved physics problems and five additional solved computer-science problems gives the 360 reported here. The primary total is 134 under either convention.

\section{Data Provenance and Reproduction}
\label{sec:provenance}
The simulation records are frozen at commit \texttt{f1275a9} (January 15, 2026). The primary reports are dated January 12, 2026; extended documentation is dated December 2025. The primary historical protocol reports seed 42 and Gemini 2.0 Flash. Preserved materials include registries, snapshots, aggregate action summaries, final agent reports, and selected source files. Complete model requests, random draws, and runtime-captured revisions are not part of the archive. Source-rule descriptions in this paper refer to the supplied snapshot.

The manifest records input source paths, commits, and SHA-256 hashes. The offline reconstruction counts unique papers, reconciles snapshots and fractional credits, normalizes network pairs, compares problem sets, and checks solve-step bounds. The query fixture tests the archived candidate functions on fixed inputs. These procedures operate on preserved records and do not invoke the simulator or an experimental model.

From the version directory, \verb|python arxiv/rebuild_evidence.py --check| checks the preserved numerical analysis, and \verb|python arxiv/eligibility_regression.py --check| checks the query fixture. The version directory includes the data and analysis scripts required for these operations. The minimal source ZIP contains the files needed to compile the manuscript. The revised figures are generated by \path{arxiv/revise_figures.py}, with numeric inputs recorded in \path{arxiv/revision_figure_data.json}. \path{README.txt} gives compilation and reproduction commands.

\end{document}